\documentclass[ reprint, showpacs,aps]{revtex4-1}
\usepackage{epsfig,graphicx,xcolor,amsbsy,amssymb,latexsym,amsfonts,amsmath}
\usepackage{pstricks}
\usepackage{color}
\usepackage{eurosym}
\usepackage{graphicx}
\usepackage{tikz}
\usepackage{amsmath,amssymb,amsfonts,pstricks}

\pdfoutput=1
\newcommand{\bea}{\begin{eqnarray}\displaystyle}
\newcommand{\eea}{\end{eqnarray}}

\vspace{-0.5cm}

\begin{document}

\title{ \vspace{-1.8cm} \textbf{Instability induced by exchange forces in a
2-D electron gas in a magnetic field with uniform gradient}\\
[15pt]}
\author{Pervez Hoodbhoy}
\date{\today}

\begin{abstract}
The exchange interaction is investigated theoretically for electrons
confined to a 2-D sample placed in a linearly varying magnetic field
perpendicular to the plane. Unusual and interesting behavior is predicted:
starting from zero electrons, as one adds electrons to the system the
maximum distance an electron can travel transverse to the $B_{z}=0$ line
(i.e. the system's width) increases continuously but this width will
subsequently begin shrinking at some critical number.However this collapse
will be reversed as the number crosses another critical value, which we
estimate here. For electron parameters typical for 2DEG's, the instability
could be observable at sufficiently low electron densities. A Hartree Fock
equation is derived. We also show that in an appropriate asymptotic limit
this leads to an approximately local potential. One key lesson is that the
exchange interaction is large and cannot be reasonably excluded from any
valid theoretical investigation. \newline
\end{abstract}

\maketitle

\affiliation{Department of Physics\\
		Forman Christian College\\
		Lahore, Pakistan.}

\tolerance=10000 \tolerance=10000

\subsection{Introduction}

Cooperative effects in many body systems have been investigated since the
early days of quantum mechanics with mean field theory being a critical and
often highly successful tool in the investigation of atomic, molecular, and
nuclear structure \cite{Negele}. The ``Fermi hole" coming from repulsion of
identical fermions with the same spin enables one to tackle a system that
would otherwise be intractable. A direct consequence is the exchange force
which is critically important in determining many body properties. While
many systems have been investigated in mean field theory, the particular
system to be discussed below has not. The present work is a first attempt to
uncover the behavior of this particular many body system teetering at the
edge of instability. As such it adds to the stock of existing systems that
are at least partially solvable \cite{Negele},\cite{Vignale}.

Considered here is a two dimensional electron gas (2-DEG) such as created by
using a GaAs/GaAlAs heterostructure. It is subjected to a perpendicular
magnetic field whose strength varies linearly from one edge to the other. M%
\"{u}ller \cite{Muller} carried out the first calculation of free electrons
with levels filled up to $\sim 14$ meV $\ $and excitation levels $n\gtrsim $ 
$30$. The sample boundary promotes electrons to the next level. In contrast
there will be no sample edge in what we consider here. As such it is a far
simpler, cleaner system. Using qualitative reasoning M\"{u}ller also pointed
out that the classical electron trajectory is snake-like and weaves around
the $B_{z}=0$ line. Other authors (including the present author) \cite%
{Reijners1}\cite{Reijners2}\cite{Nogaret}\cite{Hoodbhoy}\cite{Puja} followed
up with various other calculations but none included the exchange force. As
it turns out, the omission is crucial; exchange effects are so large that
calculations not including them might need to be reassessed or redone. Why
they were omitted is obvious: calculating the exchange energy for electrons
in a uniform magnetic field is tedious but can still be found in textbooks
such as ref\cite{Vignale} where they turn out to be substantial. A similar
calculation for the non-uniform case under consideration here seems
dauntingly complicated and seems not to have been addressed in the
literature. Any insight into the role of exchange for this particular system
would therefore be of interest.

Consider the following gedanken experiment on a rectangular surface in the $%
x-y$ plane along which electrons can move with a long mean path between
collisions. The sample is placed in a $z$-directed $B$ field, $\vec{B}%
=B^{\prime }y$ $\hat{e}_{z}.$ Starting from zero, electrons are added one by
one by, for example, changing the gate voltage. The Pauli principle
restricts electrons to higher $k_{x}$ states and thus leads to an increase
of the width because each additional electron with positive momentum can be
accommodated only to the extreme left or right. If spin is excluded,
electrons move in a symmetric potential double well. But if spin is
included, the symmetry is broken although not by very much because of the
smallness of the in-medium $g$-factor. Of course, time-reversal invariance
is always broken because of the presence of the external $B$-field and so,
placing the $x$-axis along the $B_{z}=0$ line, electrons moving vertically
up/down will experience different forces. If one kept adding electrons
indefinitely, eventually the system would expand to the horizontal size of
the sample $L_{y}$. The system does not need a confining boundary wall because for low enough densities the magnetic field keeps electrons  close to the $B_{z}=0$ line. The exact relation of width to $N$ can be easily 
derived (see Eq.\ref{size} below) if the Coulomb interaction between
electrons is turned off.

Now imagine turning on the interaction. In the mean-field approximation the
many-body wavefunction is still a Slater determinant but now the orbitals
must be determined self-consistently, and then placed to lie below the Fermi
level. As in the usual electron gas calculations there is now a direct term
as well as an exchange term. The direct term in electron gas calculations is
normally assumed to be canceled by charges in the substrate below. The
same shall be assumed here.

One might expect that, except for the system size continuously increasing in
the $y$-direction, there will be only steady but no dramatic change as one
increases $N$. But, as argued here, there is an unexpected development.
Briefly: with the direct repulsive term taken care of by the substrate
electrons, the remaining exchange interaction is attractive and so seeks to
inhibit further expansion. Depending on the size of in-medium constants, it
can induce instability and ultimately cause the system to collapse. In the
asymptotically valid analysis performed in this paper it is not possible to
determine the critical density; for this the Hartee-Fock equation derived
below, together with a constraint to be explicated, will have to be
numerically solved. Surprisingly, as one adds still more electrons, the
collapsed system is eventually forced to resume its outward expansion. The
asymptotic analysis provided here does say, at least approximately, what
this second critical density will be. As such it provides insight into the
behavior of a complex system revealing a somewhat unexpected dependence of
the critical density upon the sample vertical length $L_{x}$. It also shows
independence from the strength of the Zeeman coupling provided it is not
zero. The expected behavior is displayed, albeit only schematically, in
Fig.1. 
\begin{figure}[h]
\includegraphics[width=8cm, height=5cm]{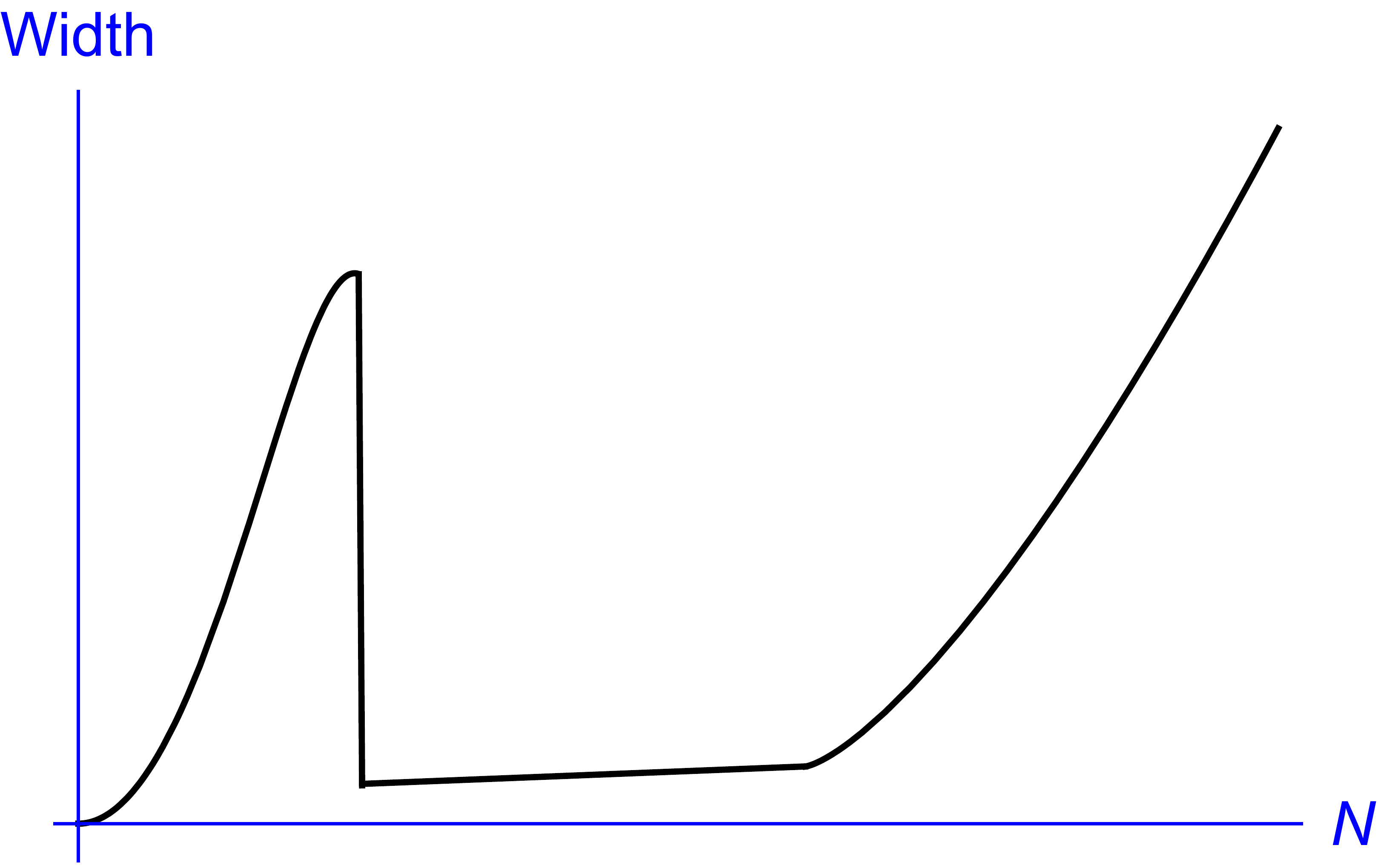} \centering
\caption{Schematic predicted behaviour of system width, i.e. length in the
direction perpendicular to the $B_{z}=0$ line, as a function of electron
number. The first critical number requires a full solution of the Hartree
Fock equation but the asymptotic analysis leading to Eq.63 yields an
estimate for the second one.}
\end{figure}
Only a full solution of the Hartree-Fock equations can reveal the true
behavior but this must be deferred to some future time.

\section{\ Preliminaries}

The starting point is the 2-D Schr\"{o}dinger equation describing free
electrons confined within a rectangular $(L_{x},L_{y})$ sample. A $\hat{z}$%
-directed magnetic field increases linearly with $y$, $\ \vec{B}%
(y)=(0,0,yB^{\prime })$.\ The origin of coordinates is taken at the sample's
centre. With the inclusion of a Zeeman term the Hamiltonian is, 
\begin{equation}
H=\frac{1}{2m}\left( \mathbf{p}-\frac{e}{c}\mathbf{A}\right) ^{2}-\vec{\mu}%
\cdot \vec{B}.  \label{Ham}
\end{equation}%
The gauge potential is chosen to be independent of $x$, 
\begin{equation}
\mathbf{A=}-\hat{x}\frac{1}{2}y^{2}B^{\prime }.
\end{equation}%
Since the Hamiltonian is invariant under translations of $x$, we have a
plane wave solution, $\psi _{k}(x,y)=\frac{1}{\sqrt{L_{x}}}%
e^{-ik_{x}x}\varphi _{k_{x}}(y)$. The quantity $k_{x}$ is the eigenvalue of $%
\hat{p}_{x}=$ $\frac{\hslash }{i}\frac{\partial }{\partial x}$. Of course, $%
\hat{p}_{x}$ is not the canonical momentum operator and so $k_{x}$ is not
the physical momentum. Translational invariance in $x$ allows imposition of
periodic boundary conditions $\psi (x+L_{x},y)=\psi (x,y)$. The sum over $%
k_{x}$ is converted to an integral in the usual way, 
\begin{equation}
N=\ \sum_{k_{x}}\longrightarrow \frac{L_{x}}{2\pi }%
\int_{k_{L}}^{k_{U}}dk_{x}\ =\frac{L_{x}}{2\pi }(k_{U}-k_{L})  \label{dens}
\end{equation}%
Given relevant fundamental physical constants at this scale, together with
the magnetic field gradient, a little experimentation leads to a unique
definition of a length scale for the system, 
\begin{equation}
L_{M}=\left( \frac{2\hslash c}{eB^{\prime }}\right) ^{1/3}.
\end{equation}%
It is useful to define the dimensionless distance $\eta =y/L_{M}$ and
wavenumber $\kappa =k_{x}L_{M}$ in terms of which,%
\begin{equation}
N=\frac{1}{2\pi }\frac{L_{x}}{L_{M}}\sigma ,\text{ \ \ }\sigma =\kappa
_{U}-\kappa _{L}.
\end{equation}%
The effective potential for $y$-motion, to be inserted into the Schrodinger
equation, is slightly asymmetric for electrons with spins parallel or
antiparallel to the field,%
\begin{equation}
\text{ }V(\eta ,\kappa )=\frac{1}{2}(\eta ^{2}-\kappa )^{2}\mp \frac{\lambda 
}{2}\eta \,.  \label{pot}
\end{equation}%
Define $\alpha ^{2}=\kappa $ for $\kappa >0$ and $\alpha ^{2}=-\kappa $ for $%
\kappa <0$. The sign of $\kappa $\ is crucial for determining the behavior
of the eigenfunctions. \ For $\kappa >0$ there are minima of the potential
located at $\eta =\pm \alpha $. On the other hand, for $\kappa <0$ the two
minima coalesce at $\eta =0.$\ In both cases, for very large $\eta $ there
is a quartic confining potential. 
\begin{figure}[h]
\includegraphics[width=8cm, height=5cm]{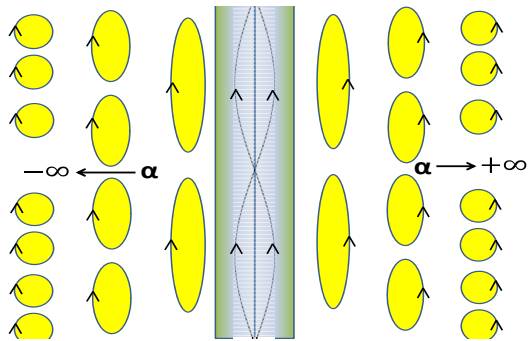} \centering
\caption{Schematic separation of regions for $\protect\kappa >0$. In the
center is the snake pit containing classical open (weaving) orbits. To the
right ($\protect\alpha >0$) and left ($\protect\alpha <0$) electron orbits
gradually close and asymptotically approach circular Landau orbits. For $%
\protect\kappa <0$ there are no closed orbits.}
\end{figure}
For a typical gradient of 1 Gauss per $\mathring{A}$, $L_{M}\approx 1096$ $%
\mathring{A}$ that is attained in physical situations, and taking the
effective electron mass $m^{\ast }\approx 0.068m_{e}$, the energy scale is
set by $\varepsilon _{0}$,%
\begin{equation}
\text{ }\varepsilon _{0}=\frac{\hslash ^{2}}{m^{\ast }L_{M}^{2}}\approx
93.25\times 10^{-6}eV.
\end{equation}%
With the in-medium electron g-factor, the dimensionless Zeeman coupling
constant $\lambda $ is, 
\begin{equation}
\lambda =g^{\ast }\frac{m^{\ast }}{m}\approx 0.0272.  \label{lam}
\end{equation}%
The Zeeman term is negligible for most purposes. But the symmetry of the
double-well potential is broken in the presence of spin; only the slightest
push suffices to send electrons over to one side or the other. One can
readily work out the asymptotic wavefunction after expanding around the
right well bottom for spins aligned along $\mathbf{B}$, 
\begin{equation}
\varphi _{\alpha }=N\exp \left[ -\alpha \sqrt{1-\frac{3\lambda }{4\alpha ^{3}%
}}\left( \eta -\alpha +\frac{\lambda }{4\alpha ^{2}}\right) ^{2}\right] .
\end{equation}%
\begin{figure}[h]
\includegraphics[width=9cm, height=6cm]{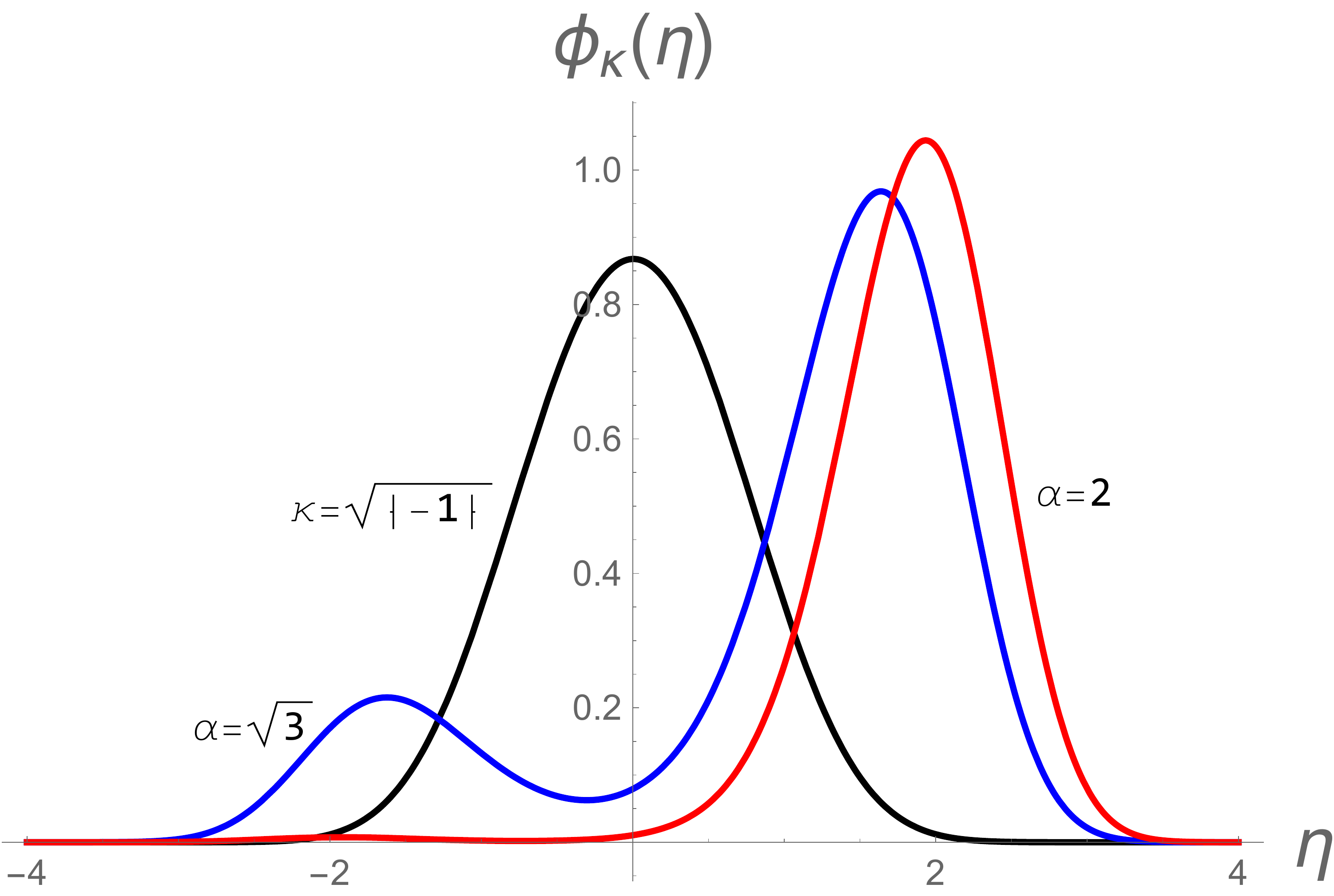} \centering
\caption{Rapid approach to asymptotia. Exact (numerical) solutions of the
one-body equation with quartic term included. Convergence to asymptotic
solution is achieved after only 2-3 magnetic lengths. Note also the strong
influence which even the tiny Zeeman term exerts; without it the ground
state wavefunction would be symmetrical at large $\protect\alpha$.}
\end{figure}
For purposes of analysis we separate three crucial regions of $k_{x}$-space:
the \textquotedblleft snake-pit" $S$ at the centre, and left/right
asymptotia,%
\begin{equation}
\kappa _{a}<L<\kappa _{b},\text{ \ \ \ }\kappa _{b}<S<\kappa _{B}\text{, \ \
\ }\kappa _{B}<R<\kappa _{A}\text{\ .}
\end{equation}%
$S$ extends roughly 2-3 magnetic length units on either side. The classical
orbits inside $S$\ are open but they close in some complicated way as one
moves outward towards $R$. For large $\kappa _{A}$, they tend towards
becoming circles because the $B$ field varies less and less over the size of
the orbit as compared to the value at the center. Quantum mechanically they
eventually become $n=0$ Landau states. Fig.2 displays a numerical
diagonalization of the Hamiltonian with the potential Eq.\ref{pot}. It shows
that the approach to right asymptotia is extremely fast. With $\lambda =0$, $%
\varphi _{\kappa }(\eta )$ would be perfectly symmetric. Notwithstanding the
tiny asymmetry of the double well potential (because of the smallness of $%
\lambda $ in Eq.\ref{lam}), we see that almost the entire wavefunction has
moved to the right for $\kappa =4$, i.e. $\alpha =2.$ For $\kappa <<0$, the
situation is still simpler for the lowest quantum state. From being only
approximately Gaussian at $\kappa =0$, it becomes nearly perfect Gaussian as 
$\kappa $ moves further to the left. We see that non-Gaussian behavior is
confined to near $S.$ Fortunately for the asymptotic analysis to be
presented below, its contribution will be negligible for large $N$.

\section{Exchange Interaction}

The starting point is the expression for the exchange energy,%
\begin{equation}
E_{ex}=-\frac{e^{2}}{2\varepsilon _{D}}\int d^{2}rd^{2}r^{\prime }\frac{\rho
(\mathbf{r,r}^{\prime })\rho (\mathbf{r}^{\prime }\mathbf{,r})}{\left\vert 
\mathbf{r}-\mathbf{r}^{\prime }\right\vert },  \label{Start}
\end{equation}%
where,%
\begin{eqnarray}
\rho (\mathbf{r,r}^{\prime }) &=&\sum_{k_{x}}\psi _{k_{x}}^{\ast }(\mathbf{r)%
}\psi _{k_{x}}(\mathbf{r}^{\prime }\mathbf{)}  \label{denmat} \\
&=&\frac{1}{2\pi }\int_{k_{L}}^{k_{U}}dk_{x}e^{ik_{x}(x-x^{\prime })}\rho
_{k_{x}}(y\mathbf{,}y^{\prime }).
\end{eqnarray}%
The electron density is invariant in the $\hat{x}$\ direction and so
requires finite integration limits but in the $\hat{y}$ direction it tails
off exponentially before reaching the sample edges. With integration regions
explicitly indicated, the exchange integral becomes,%
\begin{eqnarray}
E_{ex} &=&-\frac{e^{2}}{2\varepsilon _{D}}\int d^{2}rd^{2}r^{\prime }\frac{%
\rho (\mathbf{r,r}^{\prime })\rho \mathbf{(r}^{\prime }\mathbf{,r})}{\sqrt{%
\mathbf{(}x\mathbf{-}x^{\prime })^{2}+\mathbf{(}y\mathbf{-}y^{\prime })^{2}}}
\\
&=&-\frac{e^{2}}{2\varepsilon _{D}}\int_{-\frac{1}{2}L_{x}}^{\frac{1}{2}%
L_{x}}dxdx^{\prime }\int_{-\infty }^{\infty }dydy^{\prime }  \notag \\
&&\times \frac{e^{i(k-k^{\prime })(x-x^{\prime })}}{\sqrt{\mathbf{(}x\mathbf{%
-}x^{\prime })^{2}+\mathbf{(}y\mathbf{-}y^{\prime })^{2}}}\rho _{k_{x}}(y%
\mathbf{,}y^{\prime })\rho _{k_{x}}(y^{\prime }\mathbf{,}y).
\end{eqnarray}%
Normalizing the energy below in units of $\frac{\hslash ^{2}}{m^{\ast
}L_{M}^{2}}$ and converting all quantities in to corresponding dimensionless
variables, the energy functional becomes, 
\begin{eqnarray}
\mathcal{E[}\varphi _{\kappa }^{\ast },\varphi _{\kappa }] &=&\int_{\kappa
_{L}}^{\kappa _{U}}d\kappa \int_{-\infty }^{\infty }d\eta \varphi _{\kappa
}^{\ast }\left( -\frac{1}{2}\frac{d^{2}}{d\eta ^{2}}+V\right) \varphi
_{\kappa }  \notag \\
&&-\mu \int_{\kappa _{L}}^{\kappa _{U}}d\kappa d\kappa ^{\prime
}\int_{-l_{x}/2}^{l_{x}/2}d\xi d\xi ^{\prime }\int_{-\infty }^{\infty }d\eta
d\eta ^{\prime }  \notag \\
&&\times \rho _{\kappa }(\eta \mathbf{,}\eta ^{\prime })\rho _{\kappa
^{\prime }}(\eta ^{\prime }\mathbf{,}\eta )\frac{e^{i(\kappa -\kappa
^{\prime })(\xi \mathbf{-}\xi ^{\prime })}}{\sqrt{\mathbf{(}\xi \mathbf{-}%
\xi ^{\prime })^{2}+\mathbf{(}\eta \mathbf{-}\eta ^{\prime })^{2}}},
\label{ef}
\end{eqnarray}%
where the sample's length (in units of magnetic length) is $%
l_{x}=L_{x}/L_{M} $ and $\mu $ is the exchange coupling,%
\begin{equation}
\mu =\frac{e^{2}m^{\ast }}{8\pi ^{2}\varepsilon _{D}\hslash ^{2}}L_{M}=\frac{%
1}{8\pi ^{2}}\frac{L_{M}}{a_{B}^{\ast }},
\end{equation}%
with $a_{B}^{\ast }=\varepsilon _{D}\hslash ^{2}/e^{2}m^{\ast }$ being the
in-medium Bohr radius. For the typical values discussed above, $\mu \approx $
$0.137$. To get the equation of motion, Eq.\ref{ef} must be minimized with
respect to $\varphi _{\kappa }^{\ast }(\eta )$. Adding in a Lagrange
multiplier constraint to keep $\varphi $ normalized gives the equation of
motion,%
\begin{eqnarray}
\left( -\frac{1}{2}\frac{d^{2}}{d\eta ^{2}}+V(\eta )-\frac{\lambda }{2}\eta
\right) \varphi= \int_{\kappa _{L}}^{\kappa _{U}}d\kappa ^{\prime
}\int_{-\infty }^{\infty }d\eta ^{\prime }  \notag \\
\times \int_{-l_{x}/2}^{l_{x}/2}d\xi d\xi ^{\prime }\frac{e^{i(\kappa
-\kappa ^{\prime })(\xi \mathbf{-}\xi ^{\prime })}}{\sqrt{\mathbf{(}\xi 
\mathbf{-}\xi ^{\prime })^{2}+\mathbf{(}\eta \mathbf{-}\eta ^{\prime })^{2}}}%
\rho _{\kappa ^{\prime }}(\eta ^{\prime }\mathbf{,}\eta )\varphi _{\kappa
}(\eta ^{\prime })  \notag \\
=\varepsilon _{HF}\varphi _{\kappa }(\eta ).
\end{eqnarray}

The integration limits must be determined separately. For this we must keep
fixed the electron number while minimizing the total energy function which
takes the generic form, 
\begin{equation}
\mathcal{E[}\varphi _{\kappa }^{\ast },\varphi _{\kappa }]=\int_{\kappa
_{1}}^{\kappa _{2}}\varepsilon (\kappa )d\kappa -\mu \int_{\kappa
_{1}}^{\kappa _{2}}\int_{\kappa _{1}}^{\kappa _{2}}d\kappa d\kappa ^{\prime
}g(\kappa ,\kappa ^{\prime }).
\end{equation}%
Using a Lagrange multiplier to enforce the number constraint $\kappa
_{2}-\kappa _{1}=\sigma $ gives the additional condition, 
\begin{equation}
\varepsilon (\kappa _{2})-\varepsilon (\kappa _{1})-2\mu \int_{\text{$\kappa 
$}_{1}}^{\text{$\kappa $}_{2}}d\text{$\kappa $}\left[ g(\text{$\kappa $}_{2},%
\text{$\kappa $})-g(\text{$\kappa $}_{1},\text{$\kappa $})\right] \,=0.
\label{con}
\end{equation}%
This is an infinite set of coupled non-linear integrodifferential equations
that must be brought into some tractable form. To solve this numerically -
by iteration of course - a basis set of functions will be needed, the choice
of which will be quite crucial. One clearly needs to develop intuition if
the system is to be solved numerically. The goal here is to explore whether
some analytical results can be derived and used to illuminate these fairly
opaque equations.

\section{Effective Potential}

For systems that can be presumed to be infinite, translational invariance
can make calculation of fermion exchange effects tractable because the
simplicity of the single particle wavefunctions is maintained even in the
presence of two-body interactions. Hence, upon making a suitable gauge
choice, first order exchange corrections can be calculated for electron
Landau levels in a uniform magnetic field. But in the system under
consideration here, no translational invariance exists perpendicular to the
axis of zero $B$. Hence if one starts from a solution of the single particle
Schrodinger equation where the potential experienced by an electron owes
entirely to the external magnetic field, one expects that turning on the
Coulomb interaction between electrons would drastically change the single
particle wavefunctions. This would be especially true in a system where the
Coulomb exchange energy can be many times larger than the energy originating
from kinetic and confining terms in the single particle Hamiltonian, as
indeed does happen for the system under consideration. It was therefore
interesting to see the emergence of simplicity in an asymptotic limit.

First consider large positive $\kappa =\alpha ^{2}$, i.e. far to the right
of the snake pit and only electrons in the range $\alpha _{B}<$\ $\alpha <$\ 
$\alpha _{A}$.\ \ Interactions with electrons with the $S$ and $L$ regions
are excluded; they are in fact exponentially suppressed. For spins aligned
along the magnetic field the equation determining $\varphi _{\alpha }$ is, 
\begin{eqnarray}
&&\left( -\frac{1}{2}\frac{d^{2}}{d\eta ^{2}}+V\right) \varphi _{\alpha
}(\eta )  \notag \\
&=&4\mu \int_{\alpha _{B}}^{\alpha _{A}}\frac{\alpha ^{\prime }}{\alpha }%
d\alpha ^{\prime }\int_{-\infty }^{\infty }d\eta ^{\prime
}\int_{-l_{x}/2}^{l_{x}/2}d\xi d\xi ^{\prime }  \notag \\
&&\times \rho _{\alpha ^{\prime }}(\eta ^{\prime }\mathbf{,}\eta )\varphi
_{\alpha }(\eta ^{\prime })\frac{e^{i(\alpha ^{2}-\alpha ^{2\prime })(\xi 
\mathbf{-}\xi ^{\prime })}}{\sqrt{\mathbf{(}\xi \mathbf{-}\xi ^{\prime
})^{2}+\mathbf{(}\eta \mathbf{-}\eta ^{\prime })^{2}}}.  \label{HFEQ}
\end{eqnarray}%
In principle every electron between $\alpha _{B}$ and$\ \alpha _{A}$
interacts with every other one, a manifestation of the non-local interaction
of the exchange potential. However, in the $\alpha \rightarrow \infty $ a
series of approximations reduces the above to a relatively simple local
form. Qualitatively there are two reasons for this. First, at large $\alpha $%
\ (in the absence of exchange), the free wavefunctions are narrowly peaked
Gaussians. Thus an electron located at $\alpha $ will have exponentially
small overlap with another at $\alpha ^{\prime }$\ unless the two points are
close to each other. Hence what is non-local can hopefully be modeled with a
local potential that emerges naturally from Eq.\ref{HFEQ} \ in some
approximate way. Second, the exchange interaction in 2-D is attractive and
peaks strongly at $\alpha =\alpha ^{\prime }$ thus encouraging electrons to
come closer to each other. These qualitative considerations will be made
quantitative below.

As a first step, expand the potential about the well bottom located to the
right at $\alpha -\lambda /8\alpha ^{2}+$ $O(\lambda ^{2}).$ The smallness
of $\lambda $ means the well bottom can be safely assumed to be at $\alpha $
for large positive $\alpha $. After changing variables to $w$, 
\begin{equation}
w=\sqrt{\alpha }(\eta -\alpha ),\text{ \ }\beta =\alpha -\alpha ^{\prime }.
\end{equation}%
Eq.\ref{HFEQ} can be reexpressed as,%
\begin{eqnarray}
&&\left( -\frac{1}{2}\frac{d^{2}}{dw^{2}}+2w^{2}-\varepsilon _{HF}\right)
\varphi _{\alpha }(w)  \notag \\
&=&\mu \int_{\alpha -\alpha _{A}}^{\alpha -\alpha _{B}}d\beta h_{\alpha
}(\beta )\varphi _{\alpha -\beta }(w),  \label{H1}
\end{eqnarray}%
where,%
\begin{eqnarray}
h_{\alpha }(\beta ) &=&4\frac{\alpha ^{\prime }}{\alpha }%
\int_{-l_{x}/2}^{l_{x}/2}d\xi d\xi ^{\prime }\int_{-\infty }^{\infty }d\eta
^{\prime }  \notag \\
&&\times \frac{\varphi _{\alpha ^{\prime }}^{\ast }(\eta ^{\prime })\varphi
_{\alpha }(\eta ^{\prime })e^{i(\alpha ^{2}-\alpha ^{2\prime })(\xi \mathbf{-%
}\xi ^{\prime })}}{\sqrt{\mathbf{(}\xi \mathbf{-}\xi ^{\prime })^{2}+\mathbf{%
(}\eta \mathbf{-}\eta ^{\prime })^{2}}}  \notag \\
&=&8\frac{\alpha ^{\prime }}{\alpha }\int_{0}^{l_{x}}dt\int_{0}^{t}ds\int_{-%
\infty }^{\infty }du  \notag \\
&&\times \frac{\varphi _{\alpha -\beta }^{\ast }(\eta ^{\prime })\varphi
_{\alpha }(\eta ^{\prime })}{\sqrt{s^{2}+u^{2}}}e^{i(2\alpha -\beta )\beta s}
\label{H2}
\end{eqnarray}%
In going from Eq.\ref{H1} to Eq.\ref{H2}, new variables $t,s,u$ have been
defined, 
\begin{equation}
\xi =t+\frac{1}{2}s,\ \ \xi ^{\prime }=t-\frac{1}{2}s,\text{ \ }\eta \mathbf{%
\ -}\eta ^{\prime }\ =u.
\end{equation}%
and the change in the shape of the ($\xi ,\xi ^{\prime })$ integration plane
has been included. 
\begin{equation}
\int_{-l_{x}/2}^{l_{x}/2}d\xi d\xi ^{\prime }f(\left\vert \xi -\xi ^{\prime
}\right\vert )\rightarrow 2\int_{0}^{l_{x}}dt\int_{0}^{t}dsf(\left\vert
s\right\vert ).
\end{equation}%
The integral over $\eta ^{\prime } $or, equivalently over $u$, can be
extended to infinity if the electrons are assumed to form a self-binding
system and $\varphi (\pm \infty )=0$. However, because of translational
invariance, one cannot assume the same here in $x$. This will, as we shall
see, has profound consequences.

At this point we shall assume that the integrand in Eq.\ref{H2} is
significant only when $\beta <<\alpha $, i.e. two narrowly peaked
wavefunctions must nearly coincide to give a non-zero contribution. This
will lead to consistent results that can be verified after the calculation
is performed. As a first guess we use the lowest order solution of the free $%
\mu =0$ equation which, at leading order, yields for the product of two
wavefunctions, 
\begin{equation}
\varphi _{\alpha -\beta }^{(0)}(\eta ^{\prime })\varphi _{\alpha
}^{(0)}(\eta ^{\prime })=\sqrt{\frac{2\alpha }{\pi }}e^{-\frac{1}{2}\alpha
\beta ^{2}}e^{-2\alpha \left( u+\frac{w}{\sqrt{\alpha }}-\frac{1}{2}\beta
\right) ^{2}}.
\end{equation}%
The integrals in Eq.\ref{H2} cannot be done exactly. Since $\alpha $\ is a
large parameter one can take recourse to the method of stationary phase,
followed by steepest descent. Transforming to polar coordinates, 
\begin{equation}
u=\rho \cos \theta ,\text{ \ }s=\rho \sin \theta ,
\end{equation}%
with $0<\theta <2\pi $ and $0<\frac{t}{\left\vert \sin \theta \right\vert }%
<\rho <\infty $ gives,%
\begin{eqnarray}
h_{\alpha }(\beta ) &=&8\int_{0}^{l_{x}}dt\int_{0}^{t}ds\int_{-\infty
}^{\infty }du  \notag \\
&&\times \frac{\varphi _{\alpha -\beta }^{(0)\ast }(\eta ^{\prime })\varphi
_{\alpha }^{(0)}(\eta ^{\prime })}{\sqrt{s^{2}+u^{2}}}e^{i(2\alpha -\beta
)\beta s}  \label{c1} \\
&=&4\sqrt{2\alpha }e^{-\frac{1}{2}\alpha \beta
^{2}}\int_{0}^{l_{x}}dt\int_{0}^{t}d\rho d\theta  \notag \\
&&\times e^{-2\alpha \left( \rho \cos \theta +\frac{w}{\sqrt{\alpha }}-\frac{%
1}{2}\beta \right) ^{2}}e^{i2\alpha \beta \rho \sin \theta }  \label{c2} \\
&=&32\sqrt{2}\pi e^{-w^{2}}e^{-\alpha \left( \beta -\frac{w}{\sqrt{\alpha }}%
\right) ^{2}}  \notag \\
&&\times \int_{0}^{l_{x}}dt\int_{0}^{t}d\rho \frac{\sin (2\alpha \beta \rho
+\pi /4)}{\sqrt{\beta \rho }}.  \label{c3}
\end{eqnarray}%
\qquad In going from Eq.\ref{c1} to Eq.\ref{c2} just the leading term have
been kept, and in going from Eq.\ref{c2} to Eq.\ref{c3} it was recognized
that the phase becomes stationary at $\theta =\pi /2,3\pi /2$. Anticipating
that $\beta $ will also be integrated upon later, and noting the smooth
behavior of the remaining integrals, we can further simplify $h_{\alpha
}(\beta )$ by replacing $\beta \rightarrow w/\sqrt{\alpha }$ in the
integrand, \ 
\begin{eqnarray}
h_{\alpha }(\beta ) &=&32\sqrt{2\alpha }\pi \frac{e^{-w^{2}}}{\sqrt{%
\left\vert w\right\vert }}e^{-\alpha \left( \beta -\frac{w}{\sqrt{\alpha }}%
\right) ^{2}}  \notag \\
&&\times \int_{0}^{l_{x}}dt\int_{-t}^{t}d\rho \frac{\sin (2\alpha \beta \rho
+\pi /4)}{\sqrt{\beta \rho }}.
\end{eqnarray}%
The last two integrals can be performed exactly in terms of Fresnel
integrals $C(\sqrt{\alpha }l_{x}w)$ and $S(\sqrt{\alpha }l_{x}w)$ but the
results are not illuminating and will not be displayed here. Two limiting
cases suffice to make the point below,

We can insert Eq.\ref{c3} into Eq.\ref{H1} and keep just the first term in
the expansion about $\beta =0$. The remaining integral is trivially done and
we see the promised result, an approximate effective potential determining $%
\varphi _{\alpha }$, 
\begin{equation}
\left( -\frac{1}{2}\frac{d^{2}}{dw^{2}}+2w^{2}+\mathcal{V}(w)\right) \varphi
_{\alpha }=\varepsilon _{HF}\varphi _{\alpha },
\end{equation}%
where the limiting cases for $\mathcal{V}(w)$ can easily be worked out, 
\begin{equation}
\mathcal{V}(w)=\left\{ 
\begin{array}{lr}
-\frac{32\pi ^{2}}{\sqrt{\alpha }}l_{x}\frac{e^{-w^{2}}}{\left\vert
w\right\vert } & \left\vert w\right\vert >>\frac{1}{\sqrt{\alpha }l_{x}} \\ 
\begin{array}{l}
-\frac{128}{3}\frac{\sqrt{2\pi ^{3}}}{\alpha ^{1/4}}l_{x}^{3/2} \\ 
\times (1+\frac{2}{5}\sqrt{\alpha }wl_{x}-\frac{6}{35}\alpha w^{2}l_{x}^{2})%
\end{array}
& \left\vert w\right\vert <<\frac{1}{\sqrt{\alpha }l_{x}}%
\end{array}%
\right. .
\end{equation}%
Observe that: a)$\mathcal{V}(w)$ has a much shorter range relative to $%
2w^{2} $and so electrons confined by the potential well produced by the
magnetic field are unaffected at longer distances, b)$\mathcal{V}(w)$ is
attractive and has the effect of further narrowing the wavefunction, c)$%
\mathcal{V}(w)\rightarrow 0$ as $\alpha \rightarrow \infty $ and so the
exchange interaction vanishes asymptotically, d)$\mathcal{V}(w)$ depends on $%
l_{x}$, i.e. the length of the sample in the $x$ direction as measured in
magnetic length units. This last point is somewhat surprising. We have
argued for near locality in $y$ but linear dependence upon $l_{x}$\ shows
strong non-locality in $x$. Electrons at different $x$ positions are
definitely communicating much more with each other than with those which are
located perpendicular to the $B=0$ line. At one level this can be understood
from the $x$-independence of the electron density which follows from
translational invariance.

As a qualitative confirmation, one can insert a trial Gaussian $\psi
=Ne^{-cw^{2}}$with $c$ a variational parameter into the first order
perturbation energy, 
\begin{equation}
\varepsilon (c)=\int_{-\infty }^{\infty }\left( \frac{1}{2}\psi ^{\prime
2}+\psi (2w^{2}+\mathcal{V})\psi \right) dw.
\end{equation}%
Unfortunately the integrations are analytically too complex to be useful,
but for typical parameter values numerical integration can be readily done.
The essential point is that $\varepsilon (c)$ achieves a minimum at $%
c=c_{\min }>1$ for $\alpha >>1$\ and that $c\rightarrow 1$ as $\alpha
\rightarrow \infty $. This tells us that the wavefuncton gets even more
peaked at finite $\alpha $ and that the small $\beta $\ assumption made
earlier was indeed valid.

So far we have concentrated upon the right asymptotic region. Similar
conclusions will be drawn for the left asymptotic region: one again recovers
the free solution for $\kappa \rightarrow -\infty $, the exchange potential
leads to narrowing of wavefunctions, and for large enough $l_{x}$ the
exchange energy is again proportional to $l_{x}$.\ However these conclusions
will be based upon the analysis presented in the next section where the
reasoning will take into account the very different physics: there is only a
single well for negative $\kappa $ instead of two wells for positive $\kappa
.$

To conclude this section: the exchange term involves a six dimensional
integral that was dealt with here in a limiting case only. The results
achieved will, however, be useful in attempting a full and unconstrained
self-consistent calculation.
\section{Asymptotic Energy}
Armed with the knowledge that the asymptotic solution is the $\mu =0$\
solution - and that this will be approached for sufficiently large $\alpha $%
\ - we now calculate the action to leading order in the right and left
asymptotic regions.

\subsection{Right Region}

The energy functional in the right asymptotic region is: 
\begin{eqnarray}
\mathcal{E}^{R} &=&\int_{\alpha _{B}}^{\alpha _{A}}2\alpha d\alpha
\int_{-\infty }^{\infty }d\eta  \label{ee1} \\
&&\times \varphi _{\alpha }^{(0)}\left( {\small -}\frac{1}{2}\frac{d^{2}}{%
d\eta ^{2}}+2\alpha ^{2}(\eta -\alpha )^{2}\right) \varphi _{\alpha }^{(0)}
\\
\mathcal{E}_{ex}^{RR} &=&-\mu \int_{\alpha _{B}}^{\alpha _{A}}4\alpha \alpha
^{\prime }d\alpha d\alpha ^{\prime }\int_{-l_{x}/2}^{l_{x}/2}d\xi d\xi
^{\prime }\int_{-\infty }^{\infty }d\eta d\eta ^{\prime }  \notag \\
&&\times \rho _{\alpha }^{(0)}(\eta \mathbf{,}\eta ^{\prime })\rho _{\alpha
}^{(0)}(\eta ^{\prime }\mathbf{,}\eta )\frac{e^{i(\alpha ^{2}-\alpha
^{2\prime })(\xi \mathbf{-}\xi ^{\prime })}}{\sqrt{\mathbf{(}\xi \mathbf{-}%
\xi ^{\prime })^{2}+\mathbf{(}\eta \mathbf{-}\eta ^{\prime })^{2}}}.
\label{EE22}
\end{eqnarray}%
Here $\varphi _{\alpha }^{(0)}(\eta )=N\exp [-\alpha (\eta -\alpha )]$ is
the free solution and $\mathcal{E}^{R}$ is trivially calculated, 
\begin{equation}
\mathcal{E}^{R}=\frac{2}{3}(\alpha _{_{U}}^{3}-\alpha _{_{L}}^{3}).
\label{ee3}
\end{equation}%
However, even with the simple displaced Gaussian, the six-dimensional
integral is non-trivial because only two of the six integration have limits
that can be pushed off to infinity. However a remarkably simple analytic
result can be extracted in the large $\alpha $ limit, $1<<\alpha _{B}<\alpha
<$ $\alpha _{A}$. It is displayed below in Eq.\ref{ellip}. Arriving at the
result will need a sequence of steps beginning with a transformation to more
appropriate coordinates: 
\begin{eqnarray}
\xi &=&t+\frac{1}{2}s,\ \ \xi ^{\prime }=t-\frac{1}{2}s,\ \eta =\text{v}+%
\frac{1}{2}u,\text{ }  \label{t1} \\
\eta ^{\prime } &=&\text{v}-\frac{1}{2}u,\text{ \ }\alpha =\gamma +\frac{1}{2%
}\beta ,\text{ \ }\alpha ^{\prime }=\gamma -\frac{1}{2}\beta .
\end{eqnarray}%
Next, we suitably arrange terms in the integrand in Eq.\ref{EE22} and
perform the integrals over $u$,v (whose integration limits extend to
infinity in both directions). After some algebra and using $\gamma >>\beta $%
, four integrations remain:%
\begin{eqnarray}
\mathcal{E}_{ex}^{RR} &=&-\mu \int d\gamma d\beta \int dtds\frac{4}{\sqrt{%
\pi }}\gamma ^{5/2}  \notag \\
&&\times e^{-\gamma (\beta ^{2}-\frac{1}{2}s^{2})+i2\gamma \beta s}K_{0}(%
\tfrac{1}{2}\gamma s^{2}).
\end{eqnarray}%
$K_{0}$ is the modified Bessel function. For any even function $g(\gamma
,\beta )=g(\gamma ,-\beta )$ integrated over the $\alpha ,\alpha ^{\prime }$
integration region as indicated in Eq.\ref{EE22} one can readily show, 
\begin{eqnarray}
\int d\gamma d\beta g(\gamma ,\beta )\rightarrow \frac{1}{2}\int_{0}^{\alpha
_{A}-\alpha _{B}}d\gamma \int_{-\gamma }^{\gamma }d\beta &&  \notag \\
\times \left[ g(\frac{1}{2}\gamma +\alpha _{B},\beta )+g(-\frac{1}{2}\gamma
+\alpha _{A},\beta )\right] &&
\end{eqnarray}%
Consider now the integrals are over $\beta ,s.$ The $\beta $ integration may
be safely extended to infinity since the only support comes from the $\beta
\approx 0$ region but the $s$ integration limits are finite.\ Transform to
polar coordinates $\beta =\rho \cos \theta ,$ \ $s=\rho \sin \theta $ and $%
0<\theta <2\pi $ and $0<\frac{t}{\left\vert \sin \theta \right\vert }<\rho
<\infty .$ In this domain the phase is stationary at $\theta =\pm \frac{\pi 
}{4},\pm \frac{3\pi }{4}.$ Adding contributions in the $\theta $ integration
from all 4 points yields, 
\begin{eqnarray}
\int d\beta ds &\rightarrow &\frac{4}{\sqrt{\pi }}\gamma ^{5/2}\int d\rho
d\theta e^{-\frac{1}{4}\gamma \rho ^{2}(1+3\cos 2\theta )}  \notag \\
&&\times e^{i\gamma \rho ^{2}\sin 2\theta }K_{0}(\tfrac{1}{2}\gamma \rho
^{2}\sin ^{2}\theta ) \\
&=&16\pi \gamma ^{2}\int_{0}^{\sqrt{2}t}d\rho e^{-\frac{1}{4}\gamma \rho
^{2}}K_{0}(\tfrac{1}{4}\gamma \rho ^{2})  \notag \\
&&\times \left( \sin \gamma \rho ^{2}+\cos \gamma \rho ^{2}\right)
\end{eqnarray}%
The above integral has no analytic form for finite $t$.\ However for large
enough $l_{x}$ and $\gamma \rightarrow \infty $\ (which is the same thing as
large $\alpha $ because $\gamma =(\alpha +\alpha ^{\prime })/2$ and both $%
\alpha ^{\prime }$s are large)\ \ we can use, 
\begin{equation}
\int_{0}^{\infty }dxe^{-(1-4i)x^{2}}K_{0}(x^{2})=\sqrt{\frac{\pi }{2}}K(2i),
\end{equation}%
where $K$ is the elliptic integral of the first kind (not to be confused
with the Bessel function $K_{0}$) and $K(\pm 2i)=1.236\pm 0.389i$ . The
remaining integrals on $t$ and $\gamma $ are now trivially done to yield the
final result for the exchange contribution of electrons in the right
asymptotic region, 
\begin{eqnarray}
\mathcal{E}_{ex}^{RR} &=&-\mu l_{x}C(\alpha _{A}-\alpha _{B})^{5/2},
\label{ellip} \\
C &=&\frac{16}{5}\sqrt{2}\pi ^{3/2}(1+i)(K(-2i)-iK(2i))  \label{ellip2} \\
&=&81.8977.  \notag
\end{eqnarray}%
Again, note that the exchange energy is proportional to $l_{x}$ for large
enough $l_{x}$. Also, $\mathcal{E}_{ex}^{RR}\sim $ $\alpha _{A}^{5/2}{}$
grows less rapidly than $\mathcal{E}_{sp}^{R}\sim \alpha _{_{A}}^{3}$ . This
will be crucial when we consider the system's stability.

\subsection{Left Region}

Next, consider the energy functional in the left asymptotic region. While
there is a small risk of confusion with the notation used above for the
right asymptotic region, we shall nevertheless continue to use below the
symbols $\alpha,\kappa $ but with a crucial change of sign. In the following 
$\alpha ^{2}=-\kappa >0$ is assumed large with $\alpha _{a}>\alpha _{b}>>1$, 
\begin{eqnarray}
\mathcal{E}^{L} &=&\int_{-\kappa _{a}}^{-\kappa _{b}}d\kappa \int_{-\infty
}^{\infty }d\eta  \label{S1} \\
&&\times \varphi _{\alpha }^{\ast }\left( -\frac{1}{2}\frac{d^{2}}{d\eta ^{2}%
}+\frac{1}{2}(\eta ^{2}+\alpha ^{2})^{2}\right) \varphi _{\alpha } \\
&{\small \approx }&\int_{\alpha _{b}}^{\alpha _{a}}2\alpha d\alpha
\int_{-\infty }^{\infty }d\eta \\
&&\times \varphi _{\alpha }^{\ast }\left( -\frac{1}{2}\frac{d^{2}}{d\eta ^{2}%
}+\eta ^{2}\alpha ^{2}+\frac{1}{2}\alpha ^{4}\right) \varphi _{\alpha }%
{\small .}
\end{eqnarray}%
In the quadratic approximation the lowest eigenfunction $\varphi _{\alpha
}^{(0)}(\eta )=Ne^{-\alpha \eta ^{2}/\sqrt{2}}$ leads to 
\begin{equation}
\mathcal{E}^{L}=\frac{1}{6}(\alpha _{a}^{3}-\alpha _{b}^{3})\left( \alpha
_{a}^{3}+\alpha _{b}^{3}+3\right) .
\end{equation}%
The Coulomb exchange energy is exactly as in Eq.\ref{EE22} with new
integration limits $\alpha _{A}\rightarrow \alpha _{a}$ and $\alpha
_{B}\rightarrow \alpha _{b}.$ However physically this is a very different
physical regime where backward moving electrons are clustered along the $%
B_{z}=0$ line. Using and using $\gamma >>\beta $ and performing the two
indicated integrations with infinite limits yields, 
\begin{eqnarray}
\mathcal{E}_{ex}^{LL} &=&-\frac{4}{\sqrt{\pi }}\mu \int d\gamma d\beta \int
dtds\gamma ^{5/2}e^{\frac{1}{2}s^{2}\gamma }  \notag  \label{exc} \\
&&\times K_{0}(\frac{1}{2}s^{2}\gamma )\cos (2\beta \gamma s)  \label{ex}
\end{eqnarray}%
All integrals in Eq.\ref{ex} have finite limits. However in the limit of
both large $\gamma $ and $l_{x}$ a careful analysis shows that the $s$%
-integration limits can be pushed to infinity, giving a result in terms of a
Meijer G hypergeometric function, 
\begin{eqnarray}
\mathcal{E}_{ex}^{LL} =-\mu \frac{1}{\sqrt{\pi }}\int_{0}^{\alpha
_{a}-\alpha _{b}}\gamma ^{3/2}d\gamma  \notag \\
\times \int_{-\gamma }^{\gamma }d\beta \left[ g(\frac{1}{2}\gamma +\alpha
_{B},\beta )+g(-\frac{1}{2}\gamma +\alpha _{A},\beta )\right] \\
g(\gamma ,\beta ) =G_{2,3}^{2,2}\left( \beta ^{2}\gamma \left\vert 
\begin{array}{c}
1,1 \\ 
\frac{1}{2},\frac{1}{2},0%
\end{array}%
\right. \right) .
\end{eqnarray}%
In the the large $\gamma $ limit, $g(\gamma ,\beta )\rightarrow $ $\pi \log
(16\beta ^{2}\gamma )+\pi \boldsymbol{\gamma_{E} }$ \ where $\boldsymbol{%
\gamma_{E} } $ \ is Euler-Mascheroni constant, $\boldsymbol{\gamma_{E} }%
=0.57721$. Carrying out the remaining integrations yields the final result
for $\alpha _{A}>\alpha _{B}>>1,$%
\begin{eqnarray}
\mathcal{E}_{ex}^{LL} &=&-\mu l_{x}\alpha _{a}^{5/2}(c_{a}+c_{b}\log \alpha
_{a})  \label{EE2} \\
c_{a} &=&\frac{4\sqrt{\pi }(-92+35\sqrt{2}+30\boldsymbol{\gamma }+240\log
2-60\coth ^{-1}\sqrt{2})}{75}  \notag \\
c_{b} &=&\frac{4}{75}\sqrt{\pi }  \notag
\end{eqnarray}%
Comparing Eq.\ref{EE2} to Eq.\ref{ellip} we note that both are proportional
to $\alpha ^{5/2}$ (albeit the $\alpha ^{\prime }s$ refer to different
physical quantities) but that there is an additional logarithmic dependence
in the negative $\kappa $ case.

One issue deserves further consideration before we move on. What justifies
using $\varphi _{\alpha }^{(0)}(\eta )=Ne^{-\alpha \eta ^{2}/\sqrt{2}}$ for
large $\alpha $ in the left asymptotic region? Instead of following the line
of argument used earlier, we shall use the variational principle. To this
end, we take as trial wavefunction $\psi \sim e^{-\alpha \xi \eta ^{2}/\sqrt{%
2}}$ with $\xi $ a variational parameter. Inserting this, and going the same
sequence of steps leading to Eq.\ref{EE2}, the energy takes the following
schematic form, 
\begin{equation}
\mathcal{E}\sim \alpha _{a}^{3}\left( d+\frac{1}{d}\right) -k\alpha
_{a}^{5/2}\sqrt{d},
\end{equation}%
where crucially $k>0$ and its precise value can be read off from Eq.\ref{EE2}%
. For sufficiently large $\alpha _{a}$ the minimum is attained at $%
d=1+k\alpha _{a}^{-1/2}/4$. Thus the Gaussian becomes progressively more
peaked as one moves from infinity inwards. This confirms for $\kappa
\rightarrow -\infty $ what we had seen earlier for $\kappa \rightarrow
+\infty .$

\section{Stability/\ Instability}

A 2-D electron gas in a linearly rising magnetic field in the absence of the
exchange force, i.e. $\mu =0$, will increase its width monotonically as
electrons are added. The condition Eq.\ref{con} gives $\varepsilon (\kappa
_{2})=\varepsilon (\kappa _{1})$ or, $\frac{1}{2}\alpha _{a}^{4}=\alpha _{A}$
subject to the number conservation constraint $\alpha _{A}^{2}+\alpha
_{a}^{2}=\sigma =2\pi L_{M}\frac{N}{L_{x}}.$ This always has a solution. But
now imagine we add electrons one by one. They will, of course, go into the
lowest quantum state and first fill up the snake pit before expanding
eventually into the left/right asymptotic regions. This determines the
\textquotedblleft Fermi surface" for the system, i.e. the placement of
single electrons into single particle states.

Now turn on the exchange interaction. Keeping $\alpha _{b}$ and $\alpha _{A}$
fixed at some low values we are interested in the limit where $\alpha
_{a},\alpha _{A}$\ are in their respective asymptotic regions. With the
cautionary note that the goal is to expose the essential physics rather than
do a quantitative calculation, we shall go so far in asymptotia that only
the extreme left and right regions are relevant. Overlaps between different
regions then become negligible. Keeping only leading terms, subject to $%
\alpha _{A}^{2}+\alpha _{a}^{2}=\sigma ,$ the question becomes whether the
approximated energy, 
\begin{equation}
\mathcal{E}=\frac{1}{6}\alpha _{a}^{6}+\frac{2}{3}\alpha _{_{A}}^{3}-\mu
l_{x}(C\alpha _{A}^{5/2}+c_{a}\alpha _{a}^{5/2})  \label{et}
\end{equation}%
has a minimum. To eliminate the constraint, put $\alpha _{a}=\sqrt{\sigma }%
\sin \theta ,$ $\alpha _{A}=\sqrt{\sigma }\cos \theta .$ Since $\alpha
_{A}>>\alpha _{a}$ the last term in Eq.\ref{et} can be dropped. Thus, 
\begin{equation}
\mathcal{E}=\frac{1}{6}\sigma ^{3}\sin ^{6}\theta +\frac{2}{3}\sigma
^{3/2}\cos ^{3}\theta -\mu l_{x}C\sigma ^{5/4}\cos ^{5/2}\theta
\end{equation}%
Minimizing this with respect to $\theta $, 
\begin{equation}
\sin ^{4}\theta =\frac{2}{\sigma ^{3/2}}\cos \theta -\frac{5}{2\sigma ^{7/4}}%
\mu l_{x}C\sqrt{\cos \theta },  \label{stab}
\end{equation}%
together with positivity of the second derivative ($C$ is defined in Eq.\ref%
{ellip2}). For $\mu =0\ $and$\ \ \sigma \rightarrow \infty $, at leading
order $\theta =2^{1/4}\sigma ^{-3/8}$ and $\mathcal{E}^{\prime \prime }$ $%
\sim 2\sigma ^{-9/2}>0.$ But for finite $\mu $ the condition for the system
to become stable is, 
\begin{equation}
1>\cos \theta >\frac{25}{32\sigma ^{1/2}}\mu ^{2}l_{x}^{2}C^{2},
\end{equation}%
or approximately that, 
\begin{eqnarray}
\sigma ^{1/4} &>&\frac{5}{4\sqrt{2}}\mu l_{x}C\text{, or,}  \notag \\
2\pi L_{M}\frac{N}{L_{x}} &>&\left( \frac{5C}{32\sqrt{2}\pi ^{2}}\frac{L_{x}%
}{a_{B}^{\ast }}\right) ^{4}.  \label{crit}
\end{eqnarray}%
With this condition $\mathcal{E}^{\prime \prime }>0$ and so stability is
assured. \textbf{\ }For the system to resume expansion upon adding more
electrons, the condition derived is $L_{M}N>c$ where $c$ is some constant
that does not depend on either $N$ or $B$. Since $L_{M}\sim B^{-1/3}$, it
follows that we must either increase $N$ or decrease $B$ to achieve the same 
$c$.

A by-product of the above analysis is that we can simply read off how the
system size would increase with $N$ if exchange was absent,%
\begin{equation}
\left\langle L_{y}\right\rangle =2^{1/4}\sigma ^{1/8}L_{M}=\left( 8\pi L_{M}%
\frac{N}{L_{x}}\right) ^{1/8}L_{M}.  \label{size}
\end{equation}%
In Eq.\ref{size} $\left\langle L_{y}\right\rangle $ is equal to half the
system's width, i.e. the distance from the $B_{z}=0$ line to the last
occupied state on the right (labeled by $\alpha _{a}$) provided: a)the
sample boundary lies even further to the right, b)the exchange interaction
is negligible $(\mu =0)$, c)$N$ is large enough so that the snake-pit is
irrelevant.

\section{Discussion}

With $L_{x}=$1cm $\approx 10^{5}L_{M}$ for a 1 Gauss/$\mathring{A}$\ \
gradient and the physical constants specified earlier, the condition from Eq.%
\ref{crit} is $\sigma ^{1/4}>9\times 10^{5}$ which corresponds to $N\sim
1.3\times 10^{10}$. To put this in context, note that in high mobility
molecular-beam-epitaxially grown GaAs-AlGa heterostructures using electron
beam lithography, electron densities are typically around $10^{11}$cm$^{-2}.$
This suggests that an experimental check may not be too difficult.

The instability discussed in this paper can be understood using some
hand-waving arguments. Far from the snake pit, each electron is in a locally
constant $B$ field and hence approaches the characteristic motion of an
electron in its lowest Landau level with a $B$ equal to that at the center
of gyration. Because of the Pauli principle, it would appear that the double
well always requires a system with more electrons to be larger. But let's
recall that that the kinetic-magnetic energy (positive) is proportional to
the one-body density matrix $\rho $ while the exchange energy (negative) is
proportional to $\rho ^{2}.$ So, loosely speaking, as one adds electrons and
increases the density, an electron acquires more neighbors. With the Pauli
principle allowing only one electron in the state labeled by $\alpha $, the
kinetic-magnetic term grows as $\alpha ^{3}\sim y^{3},$ i.e. as the cube of
the system's width. The exchange energy grows somewhat more slowly as $%
\alpha ^{5/2}$ and so, independent of the coefficient in front of it, will
eventually become sub-dominant. The catch, however, is that the
electron-electron effective coupling is large and so at some point it
becomes energetically favorable for electrons to bunch together. That leads
to collapse, i.e. the width decreases as $N$ increases. However, a further
increase of $N$ eventually causes the kinetic-magnetic term to win and thus
the system will resume expanding (until such time as it hits the boundary,
which is excluded here).

Several interesting issues could be investigated at a later time. Obviously,
the first is to set up the computational machinery for solving the
Hartree-Fock equations whose solutions alone can give a definite value for
when the first discontinuity occurs. The most suitable basis set for solving
the HF equation is to take a set of simple harmonic oscillator functions
that is not centered at $\eta =0$ but, instead, moves with the peak of the
wavefunction and so is centered \ close to $\eta =\alpha $ (see Fig.3). Of
course one must compute overlaps between wavefunctions centred at different
points and that brings in its own set of complications when calculating the
6-dim exchange integral. However a preliminary investigation shows it is
likely to work efficiently.

One also needs to understand better the dependence of the coupling on $L_{x}$
and whether the periodic boundary conditions properly model the physical
situation. Naively one should be able to push off the $x$-boundary to
infinity because the Coulomb potential falls off with distance. However it
does not fall off fast enough and so there is communication between
electrons at roughly the same horizontal distance as the electrons near the
sample's edges. The resulting coupling is linear in $L_{x}$ only in the $%
L_{x}\rightarrow \infty $ limit; at finite size there appear to be
Fresnel-like oscillations.

A separate question is what happens when the system expands to the
horizontal boundary. While fairly trivial in the absence of exchange forces,
it too could bring some surprises as the forces at the boundary push
electrons into higher Landau levels and skipping states. Finally: what
happens to gauge dependence? It is perfectly valid to work consistently
within a single gauge (as done here) and calculate physical quantities. But,
in principle, one should be able to make a different gauge choice and show
that the same physical quantities emerge.

\end{document}